\documentclass[twocolumn,showpacs,preprintnumbers,amsmath,amssymb]{revtex4}

\usepackage{graphicx}
\usepackage{dcolumn}
\usepackage{bm}

\begin{document}

\preprint{}

\title{ Controlled and secure  direct communication using GHZ state and teleportation}

\author{Ting  Gao $^{1, 2}$ }

 \affiliation
  {$^1$ Department of Mathematics, Capital Normal University, Beijing 100037, China\\
$^2$ College of Mathematics and Information Science, Hebei Normal
University, Shijiazhuang 050016, China\\}
\date{\today}
\begin{abstract}
A theoretical  scheme for controlled and  secure direct communication is proposed. The communication is based on
GHZ state and controlled quantum teleportation. After insuring the security of the quantum channel (a set of
qubits in the GHZ state), Alice encodes the secret message directly on a sequence of particle states  and
transmits them  to Bob supervised by Charlie using controlled quantum teleportation. Bob can read out the
encoded messages directly by the measurement on his qubits. In this scheme, the controlled quantum teleportation
transmits Alice's message without revealing any information to a potential eavesdropper. Because there is not a
transmission of the qubit
 carrying the secret messages between Alice and Bob in the public channel, it is
completely secure for controlled and  direct secret communication
if perfect quantum channel is used. The feature of this scheme is
that the communication between two sides depends on the agreement
of the third side.
\end{abstract}

\maketitle

Cryptography is an art  to ensure that the secret message is intelligible only for the two authorized parties of
communication and can not be altered during the transmission. It is generally believed that cryptography schemes
are only completely secure when the two communicating parties establish a shared secret key before the
transmission of a message. It is trusted that the only proven secure crypto-system is the one-time-pad scheme in
which the secret key is as long as the message. But it is difficult to distribute securely the secret key
through a classical channel. Fortunately, people did discover protocols for secure key distribution.  As shown
in a seminal paper by Bennett and Brassard in 1984 \cite {s1}, Alice and Bob can establish a shared secret key
by exchanging single qubits, physically realized by the polarization of photons, for example. The security of
this quantum key distribution is guaranteed by the principle  of quantum mechanics. Up to now there have been a
lot of theoretical quantum key distribution schemes such as in Refs. \cite {s1, s2, s3, s4, s5, s6, s7, s8, s9,
s10, s11, s12, s13, s14, s15, s16, s17, s18}.

 Recently, a novel quantum direct communication protocol has be
presented \cite {s19} that allows secure direct communication, where there is no need for establishing a shared
secret key and the message is deterministically sent through the quantum channel, but can only be decoded after
a final transmission of classical information. Bostr\"{o}m and Felbinger \cite {s20} put forward  a direct
communication scheme, the "ping-pong protocol", which also allows for deterministic communication. This protocol
can be used for the transmission of either a secret key or a plaintext message. In the latter case, the protocol
is quasi-secure, i.e. an eavesdropper is able to gain a small amount of message information before being
detected. In case of a key transmission the protocol is asymptotically secure if perfect quantum channel is
used. But it is insecure if it operated in a noisy quantum channel, as indicated  by W\'{o}jcik  \cite {s21}.
There is some probability that a part of the messages might be leaked to the eavesdropper, Eve, especially in a
noisy quantum channel, because Eve can use the intercept-resending strategy to steal some secret messages even
though Alice and Bob will find out her in the end of communication. More recently Deng et al. \cite {s22}
suggested a two-step quantum direct communication protocol using Einstein-Podolsky-Rosen pair block. It was
shown that it is provably secure. However in all these secure direct communication schemes it is necessary to
send the
  qubits carrying  secret messages in the public channel. Therefore,
  Eve can  attack the qubits  in
   transmission. Yan and Zhang \cite {s23} presented  a scheme for secure direct
   and confidential communication between Alice and Bob, using
   Einstein-Podolsky-Rosen pairs and teleportation
    \cite {s24}.   Because  there is not a transmission of  the qubits
    carrying the secret messages between Alice
and Bob in the public channel, it is completely secure for direct
secret communication if perfect quantum channel is used.

Quantum teleportation was invented by Bennett et al. \cite {s24} and developed by many authors \cite {s25, s26}.
In 2000, Zhou et al. proposed a controlled quantum teleportation scheme  \cite {s25}, where the entanglement
property of GHZ state is utilized. According to the scheme, a third side is included, so that the quantum
channel is supervised by this additional side. The signal state can not be transmitted unless all three sides
agree to cooperate.

  In this paper we design a scheme for  controlled and secure direct communication
 based on GHZ state
and controlled quantum teleportation.
 The feature of this scheme is that the communication  between two
sides depends on the agreement of the third side.

  The new protocol can be divided into two steps, one is to prepare  a set of triplets of  qubits in the
   GHZ state  (quantum channel), the other is to
  transmit messages using  controlled quantum teleportation.

{\it Preparing quantum channel} --- Suppose that Alice,  Bob and Charlie share  a set of triplets of qubits  in
 GHZ state
\begin{equation*}
|\Phi^+\rangle_{ABC}=\frac {1}{\sqrt
2}(|000\rangle-|111\rangle)_{ABC}.
\end{equation*}
 Obtaining these triplets of particles in GHZ state could have come about in many different ways; for example, Alice
, Bob or Charlie could prepare the triplets and then send one qubit of each triplet to each of the other two
persons (that is, one of them generates and shares each of the triplets with the other two people.).
 Alternatively,
 a fourth  party could prepare  an ensemble of particles in GHZ state, and ask Alice, Bob and Charlie   to  each
 take a particle ($A$, $B$, $C$, respectively) in each triplet.  Or they could have met
 a long time ago and shared them,
 storing them until the present.  Alice, Bob and Charlie  then choose  randomly a subset of qubits
 in   GHZ state,
 and do   some  appropriate tests of fidelity.  Passing the test certifies that
  they continue to hold sufficiently pure, entangled quantum states.  However,
 if tampering has occurred,
 Alice, Bob and Charlie discard  these triplets, and  a new set of qubits in GHZ
 state should be constructed
 again.

{\it Secure direct communication using controlled teleportation}
--- After insuring the security of the quantum channel (GHZ state), we begin  controlled and secure direct communication.  Suppose that Alice has a particle sequence and she
wishes to communicate information to Bob supervised by Charlie. First Alice makes his particle sequence in the
states, composed of $|+\rangle$ and $|-\rangle$, according to the message sequence. For example if the message
to be transmitted is 101001, then the sequence of particle states should be in the state
$|+\rangle|-\rangle|+\rangle|-\rangle|-\rangle|+\rangle$, i.e. $|+\rangle$ and $|-\rangle$ correspond to 1 and 0
respectively. Here
\begin{equation*}
|+\rangle=\frac {1}{\sqrt 2}(|0\rangle+|1\rangle),
~~~~~~~~~~|-\rangle=\frac {1}{\sqrt 2}(|0\rangle-|1\rangle).
\end{equation*}

Remarkably quantum entanglement of GHZ state can serve as a channel for transmission of message encoded in the
sequence of particle states. This is the process so called controlled quantum teleportation \cite {s25} which we
now describe. Suppose the quantum channel $|\Phi^+\rangle_{ABC}$ shared by particles $A$, $B$ and $C$ belong to
Alice, Bob and Charlie, respectively.  In components we write the signal state carrying secret message
\begin{equation*}
|\Psi\rangle_D=\frac {1}{\sqrt 2}(|0\rangle+b|1\rangle)_D,
\end{equation*}
where $b=1$ and $b=-1$ correspond to $|+\rangle$ and $|-\rangle$ respectively.   The quantum state of the whole
system (the four qubits) can be written as
\begin{eqnarray*}
&&~~~|\Psi\rangle_D|\Phi^+\rangle_{ABC}\\
&&=\frac {1}{\sqrt 2}(|0\rangle+b|1\rangle)_D\frac {1}{\sqrt 2}(|000\rangle-|111\rangle)_{ABC}\\
&&=\frac {1}{2}\cdot\frac {1}{\sqrt 2}(|00\rangle+|11\rangle)_{DA}\frac {1}{\sqrt
2}(|00\rangle-b|11\rangle)_{BC}
\\
&&~~+\frac {1}{2}\cdot\frac {1}{\sqrt 2}(|00\rangle-|11\rangle)_{DA}\frac {1}{\sqrt
2}(|00\rangle+b|11\rangle)_{BC}
\\
&&~~+\frac {1}{2}\cdot\frac {1}{\sqrt 2}(|01\rangle+|10\rangle)_{DA}\frac {1}{\sqrt
2}(b|00\rangle-|11\rangle)_{BC}
\\
&&~~+\frac {1}{2}\cdot\frac {1}{\sqrt 2}(|01\rangle-|10\rangle)_{DA}\frac {1}{\sqrt
2}(-b|00\rangle-|11\rangle)_{BC}.
\\
\end{eqnarray*}
Now Alice performs a Bell state measurement \cite {s27, s28} on qubits $DA$ and then  broadcasts the outcome of
her measurement. Depending on Alice's four possible  measurement outcomes $\frac {1}{\sqrt
2}(|00\rangle+|11\rangle)_{DA}, \frac {1}{\sqrt 2}(|00\rangle-|11\rangle)_{DA}, \frac {1}{\sqrt
2}(|01\rangle+|10\rangle)_{DA}$ and $\frac {1}{\sqrt 2}(|01\rangle-|10\rangle)_{DA}$, Bob and /or Charlie can
transform qubits $BC$ to  a common form:
\begin{equation*}
|\Psi\rangle_{BC}=\frac {1}{\sqrt 2}(|00\rangle+b|11\rangle)_{BC}
\end{equation*}
by the corresponding transformations  $I_B\otimes(|0\rangle\langle 0|-|1\rangle\langle 1|)_C$, $I_B\otimes I_C$,
$(|0\rangle\langle 1|+|1\rangle\langle 0|)_B\otimes(-|0\rangle\langle 1|+|1\rangle\langle 0|)_C$ and
$(-|0\rangle\langle 1|-|1\rangle\langle 0|)_B\otimes(|0\rangle\langle 1|+|1\rangle\langle 0|)_C$, respectively.

 As a matter of fact,  at this moment, neither Bob nor Charlie can
obtain the signal state $\frac {1}{\sqrt 2}(|0\rangle+b|1\rangle)$
without the cooperation of the other one.

If Charlie would like to help Bob for the quantum teleportation, he should just measure his portion of $BC$,
namely qubit $C$, on the base $\{ |+\rangle_C, |-\rangle_C\}$, and transfer the result of his measurement to Bob
via a classical channel. Here the state of qubits $BC$ can be expressed as :
\begin{eqnarray*}
&&~~~|\Psi\rangle_{BC}\\
&& =\frac {1}{\sqrt 2}(|00\rangle+b|11\rangle)_{BC}\\
&& =\frac {1}{\sqrt 2}[\frac {1}{\sqrt
2}(|0\rangle+b|1\rangle)_{B}|+\rangle_C +\frac {1}{\sqrt
2}(|0\rangle-b|1\rangle)_{B}|-\rangle_C].
\end{eqnarray*}

Once Bob has learned Charlie's result, he can "fix up" his state, recovering $|\Psi\rangle_D$, by applying an
appropriate unitary transformation. In fact, according to the two possible results $|+\rangle_C$ and
$|-\rangle_C$, Bob can perform  the corresponding transformations  $I_B$ and $(|0\rangle\langle
0|-|1\rangle\langle 1|)_B$, respectively, on qubit $B$ to obtain the signal state,
\begin{equation*}
|\Psi\rangle_B=\frac {1}{\sqrt 2}(|0\rangle+b|1\rangle)_B.
\end{equation*}
Then Bob measures  the base $\{|+\rangle, |-\rangle\}$ and reads out the  messages that Alice wants to transmit
to him.

  It is undeniable that this process of controlled quantum teleportation has similar notable features of the original
 quantum  teleportation [24] which was mentioned in [23]. For instance, the process
is entirely unaffected by any noise in the spatial environment between each other, and  the controlled
teleportation achieves perfect transmission of delicate information across a noisy environment and without even
knowing the locations of each other.  In the process Bob is left with a perfect instance of $|\Psi\rangle$ and
hence no participants can gains any further information about its identity. So in our scheme controlled quantum
teleportation transmits Alice's message without revealing any information to a potential eavesdropper, Eve, if
the quantum channel is  perfect GHZ state (perfect quantum channel).

The security  of this protocol only depends on the perfect quantum
channel (pure GHZ state). Thus as long as the quantum channel is
perfect, our scheme is absolutely reliable, deterministic and
secure.

 Of course, we should pointed out that it is necessary for testing the
security of quantum channel, since a potential eavesdropper may
obtain information as following:

(1) Eve can use the entanglement triplet in GHZ state to obtain information. Suppose that Eve has triplets of
qubits $EFG$ in the state $\frac {1}{\sqrt 2}(|000\rangle-|111\rangle)_{EFG}$. When Eve obtains particles $B$
and $C$ (the other cases are similar to this) in preparing GHZ state, she performs a  measurement  on the
particles $BCE$ using the base $\{\frac {1}{\sqrt 2}(|000\rangle-|111\rangle),  \frac {1}{\sqrt
2}(|000\rangle+|111\rangle), \frac {1}{\sqrt 2}(|001\rangle-|110\rangle), \frac {1}{\sqrt
2}(|001\rangle+|110\rangle), \frac {1}{\sqrt 2}(|010\rangle-|101\rangle), \frac {1}{\sqrt
2}(|010\rangle+|101\rangle), \frac {1}{\sqrt 2}(|100\rangle-|011\rangle), \frac {1}{\sqrt
2}(|100\rangle+|011\rangle)\}$. From the following expression
\begin{eqnarray*}
&&~~~|\Phi^+\rangle_{ABC}|\Phi^+\rangle_{EFG}\\
&& =\frac {1}{2}[\frac {1}{\sqrt 2}(|000\rangle-|111\rangle)_{BCE}\frac {1}{\sqrt
2}(|000\rangle-|111\rangle)_{AFG}\\
&&~~~+\frac {1}{\sqrt 2}(|000\rangle+|111\rangle)_{BCE}\frac {1}{\sqrt
2} (|000\rangle+|111\rangle)_{AFG}\\
&&~~~+ \frac {1}{\sqrt 2}(|001\rangle-|110\rangle)_{BCE}\frac
{1}{\sqrt 2}(-|011\rangle+|100\rangle)_{AFG}\\
&&~~~+\frac {1}{\sqrt 2}(|001\rangle+|110\rangle)_{BCE}\frac {1}{\sqrt 2} (-|011\rangle-|100\rangle)_{AFG}].
\end{eqnarray*}
we can read off the possible post-measurement states of particles  $AFG$ $\frac {1}{\sqrt
2}(|000\rangle-|111\rangle)_{AFG}$,
 $\frac {1}{\sqrt 2}(|000\rangle+|111\rangle)_{AFG}$, $\frac {1}{\sqrt 2}(-|011\rangle+|100\rangle)_{AFG}$, $\frac
{1}{\sqrt 2}(-|011\rangle-|100\rangle)_{AFG}$ depending on Eve's possible  measurement outcomes  $\frac
{1}{\sqrt 2}(|000\rangle-|111\rangle)_{BCE}$, $\frac {1}{\sqrt 2}(|000\rangle+|111\rangle)_{BCE}$, $\frac
{1}{\sqrt 2}(|001\rangle-|110\rangle)_{BCE}$ and $\frac {1}{\sqrt 2}(|001\rangle+|110\rangle)_{BCE},$
respectively. Then Eve
 transmits the particles $B$ and $C$ to Bob and Charlie respectively. Alice, Bob and Charlie proceed  as usual,
 since they do not know that there is a
potential eavesdropper intercepting and resending  their particles  if they do not test the quantum channel.
Therefore a part of messages might be leaked to Eve.

However, by testing quantum channel, Alice, Bob and Charlie can find Eve and avoid the information being leaked.
In fact after the measurement performed by Eve,  there is not any correlation between particles  $A$ and $B$
 and particles $A$ and $C$.
  So when Alice,  Bob  and Charlie perform  measurements on oneself's particle using
  the base $\{|0\rangle, |1\rangle\}$ independently, the results will be random without  any correlation.
 If this case occurred,
 they can assert that an  eavesdropper exists  and the triplets in GHZ state should be discarded.

 (2) Eve can obtain information by  coupling the qubits in  GHZ state with her probe in  preparing GHZ state. In this case
 Alice, Bob and Charlie can also test whether the quantum channel is perfect or not  by the following strategy.
 They select at random a subset of triplets of qubits in GHZ state. All three
 measure $\sigma_x$ on some of the particles at their disposal, $\sigma_y$ on the others,  and then inform each other of
 the measurement outcomes and the corresponding operators.  When two of
the friends measure $\sigma_y$ and the third measures $\sigma_x$ on a triplet, and all three of them measure
$\sigma_x$ on triplet, it just so happens that $|\Phi^+\rangle_{ABC}$ is an eigenstate of the three operator
products $\sigma_y^A\sigma_x^B\sigma_y^C$, $\sigma_y^A\sigma_y^B\sigma_x^C$,
 $\sigma_x^A\sigma_y^B\sigma_y^C$ with eigenvalue 1 and is also an eigenstate of $\sigma_x^A\sigma_x^B\sigma_x^C$
 with eigenvalue -1. (Here $\sigma_x^A$ Alice's spin, $\sigma_x^B$ operates on Bob's, etc.) Thus if Alice, Bob
 and Charlie all measure $\sigma_x$, they may obtain -1, -1, -1 or -1, 1, 1 or 1,-1, 1 or 1, 1, -1 respectively;
 if two measure $\sigma_y$ and the third measures $\sigma_x$, they may obtain 1, 1, 1 or 1, -1, -1 or -1, 1, -1
 or -1, -1, 1 respectively; only these results are possible  (i.e. these results are complete correlation.). Here
 we can say that Eve does not exist. However,
  If other outcomes appear or measurement outcomes are not complete correlation, they can affirm that a
 potential Eve exists and have coupled the triplets of qubits in GHZ state with her probe.
The reason is as follows:

  As a matter of fact, the  overall  state of the qubits of  Alice, Bob, Charlie and Eve in general form is
  \begin{eqnarray*}
&&|\Psi\rangle_{ABCE}=|000\rangle|e_{000}\rangle +
|001\rangle|e_{001}\rangle + |010\rangle|e_{010}\rangle \\
&&~~~~~~~~~~~~~~~+ |011\rangle|e_{011}\rangle +|100\rangle|e_{100}\rangle+|101\rangle|e_{101}\rangle
\\
&&~~~~~~~~~~~~~~~+|110\rangle|e_{110}\rangle+|111\rangle|e_{111}\rangle,
  \end{eqnarray*}
where $|e_{ijk}\rangle$ ($i,j,k=0,1$) is a state of Eve's particles. Suppose that  $|\Psi\rangle_{ABCE}$ is an
eigenstate of $\sigma_x^A\sigma_x^B\sigma_x^C$ with  eigenvalue -1, it must be
 \begin{eqnarray*}
&&~~~|\Psi\rangle_{ABCE}\\
&&=\frac {1}{\sqrt 2}(|000\rangle-|111\rangle)|e'_{000}\rangle +
\frac {1}{\sqrt 2}(|001\rangle-|110\rangle)|e'_{001}\rangle \\
&&~~~+\frac {1}{\sqrt 2}(|010\rangle-|101\rangle)|e'_{010}\rangle + \frac {1}{\sqrt
2}(|011\rangle-|100\rangle)|e'_{011}\rangle.
  \end{eqnarray*}
At the same time,  assume that  $|\Psi\rangle_{ABCE}$  is also   an eigenstate of
$\sigma_y^A\sigma_x^B\sigma_y^C$, $\sigma_y^A\sigma_y^B\sigma_x^C$,
 $\sigma_x^A\sigma_y^B\sigma_y^C$ with eigenvalue 1,   so $|\Psi\rangle_{ABCE}$
 must be
\begin{equation*}
|\Psi\rangle_{ABCE}=\frac {1}{\sqrt 2}(|000\rangle-|111\rangle)|e''_{000}\rangle.
\end{equation*}

From this fact we conclude  as long as $|\Psi\rangle_{ABCE}$ is the simultaneous eigenstate   of the operators
$\sigma_y^A\sigma_x^B\sigma_y^C$, $\sigma_y^A\sigma_y^B\sigma_x^C$,
 $\sigma_x^A\sigma_y^B\sigma_y^C$ and
 $\sigma_x^A\sigma_x^B\sigma_x^C$ with the  eigenvalues 1, 1, 1, and -1 respectively, there is no entanglement
 between
  Alice, Bob and Charlie's
particles and Eve's particles. So when Alice, Bob and Charlie confirm that their qubits are complete
correlation, then  Eve can not obtain any information. If the situation  is not  the case, evidently there is a
potential eavesdropper. We should abandon the quantum channel.

In one word, under any case, as long as an eavesdropper exists, we can find her and insure the security of
quantum channel to realize controlled and secure direct communication.

In summary, we give a scheme for controlled and secure  direct communication.  The communication is based on GHZ
state and controlled teleportation. After insuring the security of the quantum channel (GHZ states), Alice
encodes the secret messages directly on a sequence of particle states  and transmits them  to Bob by
teleportation supervised by Charlie. Evidently controlled teleportation transmits Alice's messages without
revealing any information to a potential eavesdropper. Bob can read out the encoded messages directly by the
measurement on his qubits. Because there is not a transmission of the qubit which carries the secret message
between Alice and Bob, it is completely secure for controlled and direct secret communication if perfect quantum
channel is used.

Teleportation has been realized in the experiments \cite {s29, s30, s31}, therefore our protocol for controlled
and  secure direct communication will be realized by  experiment easily.

\begin{acknowledgements}
 This work was supported by National Natural Science Foundation of
China under Grant No. 10271081 and Hebei Natural Science
Foundation under Grant No. 101094.
\end{acknowledgements}

\end{document}